\documentclass{article}
\usepackage[utf8]{inputenc}
\usepackage{xcolor}
\usepackage{authblk}

\title{Debris Disks} 
\author[]{Mark C. Wyatt}
\affil[]{Institute of Astronomy, University of Cambridge, Madingley Road, Cambridge, CB3 0HA, UK}
\date{}
\usepackage[numbers]{natbib}
\usepackage{graphicx}
\usepackage{geometry}
\usepackage[utf8]{inputenc}

\begin{document}
\maketitle

\section{Introduction}  
{\it Debris disk} is a catch-all term that can be used to refer to any component of a planetary
system which is not an actual planet.
In the Solar System this refers to the asteroids and comets in the Asteroid and Kuiper belts as
well as the dust and gas derived from them, such as the zodiacal cloud.
Studying the structure of extrasolar debris disks provides unique constraints on the underlying
planetary system and on the processes of planet formation and protoplanetary disk evolution.
Debris disks also have important implications for processes affecting the planets, such as impact
events that may strip an atmosphere or deliver volatiles.
Their presence or absence also has consequences both positive and negative for the detectability of
exoplanets in the system.
Here I describe five of the most important ways in which debris disks will contribute to our
understanding of extrasolar planetary systems in the coming years.

\section{Placing the Solar system in context} 
\label{sec:ss}
Debris disks were first discovered over 30 years ago from the far-infrared emission
of circumstellar dust heated by the star \citep{Aumann1984}.
This continues to be the main method of discovery and several hundred
such cold debris disks are known, with temperatures that place the dust at 10s of au from their stars (i.e., at distances comparable with our own Kuiper belt).
Most recently {\it Spitzer} and {\it Herschel} surveys have quantified the incidence of
detectable cold emission as $\sim 20$\% across the spectral type range A-K
\citep{Eiroa2013,Thureau2014,Sibthorpe2018}
and has shown that such disks decay in brightness as the stars age
\citep{Rieke2005,Wyatt2008}
in a manner that can be explained by mass loss due to collisional erosion within
a population of planetesimal belts 
\citep{Wyatt2007,Lohne2008,Gaspar2013}.
However, we are not yet able to detect true analogues to the Solar System's debris disk
\citep{Booth2009,Vitense2012}.
Thus we cannot answer how typical the Solar System is in terms of its Kuiper belt and zodiacal dust levels (i.e., where it sits amongst the $\sim 80$\% of stars without detectable debris).
This is important to place our best example of a habitable planetary system into context, especially if debris played a role in its habitability and in the evolution of life \citep[e.g.,][see also \S \ref{sec:bomb}]{Alvarez1980,Patel2015}.

While {\it Herschel} had the sensitivity to achieve detections at a few times Solar system levels
\citep{Eiroa2013}, such photometric surveys are fundamentally limited in that a Kuiper belt
analogue emits less than 1\% of the stellar flux at all wavelengths, whereas neither the stellar flux
nor the instrumental calibration are known that accurately \citep[e.g.,][]{Liseau2013}.
Essentially this has limited our current knowledge to the brightest 20\% of extrasolar Kuiper belts
and the brightest $\sim 1$\% of exozodiacal clouds \citep[i.e., analogues to the cometary dust seen in
the inner Solar system;][]{Kennedy2013}.
To detect true Kuiper belt analogues, these would have to be resolved from the
stellar emission, which requires high resolution imaging capabilities with
well characterised instrumental systematics to reach the high contrast of the disk relative to
the star, as well as techniques to mitigate background confusion at some wavelengths \citep[e.g.,
in the far-IR where there is a high density of extragalactic sources,][]{Nguyen2010}.

Recent improvements in nulling mid-IR interferometric techniques have resulted in detections of exozodiacal emission with {\it LBTI} at 30 times Solar system levels \citep{Ertel2020}.
However, this still means that it is only the brightest $\sim 20$\% of exozodi that have been detected,
and inferences about lower dust levels have to be gleaned by extrapolation.
Yet, such low levels are important not only for their astrophysical context (i.e., how the
Solar system's cometary dust levels compare with other stars), but also because this
dust is a serious hindrance in our ability to detect emission from exo-Earths and its
presence or absence must be factored in to the design of telescopes such as {\it HabEx}, {\it LUVOIR} and {\it LIFE} \citep{Roberge2012,Quanz2019}.
For now such extrapolations suggest that the median exozodi level could be as low as 3 times Solar system levels \citep{Ertel2020}, which would be good news for exo-Earth detection.
However, this remains to be confirmed with more sensitive observations and we can only confidently say this median is $<27$ times Solar System levels.

Fortunately there is a range of instrumentation coming on-line that will help progress towards the
detection of true analogues to the Solar system's debris disk, and so our understanding of
the Solar system's place within the wider population.
The {\it LBTI} techniques and instrumentation can be improved, and {\it JWST} provides high resolution and
sensitivity mid-IR imaging that is imminent.
Shortly thereafter, even higher resolution capabilities will become available with {\it ELT} and coronagraphic imaging with {\it WFIRST}.
Further in the future far-IR observatories such as {\it Spica} and the {\it Origins Space Telescope},
if approved, would have the capability of detecting true Kuiper belt analogues towards the
nearest stars \citep{Roelfsema2018}.
In the meantime, studies of the brightest 20\% of debris disks also provide a perspective on how the Solar system may have looked early in its history, before the system-wide instability that depleted the Kuiper belt \citep[][see also \S \ref{sec:arch}]{Booth2009}, or shortly after giant impact events like that which formed the Moon \citep[][see also \S \ref{sec:bomb}]{Jackson2012}.

\section{Constraints on planetary system architecture and evolutionary history}
\label{sec:arch}
The proximity of nearby stars (10s of pc) and the large size of most debris disks
(typical radii in the range 30-150\,au) means that, for the brightest disks their emission
can be spatially resolved providing a map of the dust distribution in the system.
Indeed the first debris disk was imaged around the star $\beta$ Pictoris soon after its discovery \citep{Smith1984}, though it was not until 14 years later that the next five debris disk images were taken \citep{Holland1998,Greaves1998,Jayawardhana1998}.
By now, well over 100 debris disks have been spatially resolved, at a range of wavelengths from the
optical, at which starlight scattered by the dust is seen, and where the stellar emission has to be
suppressed by coronagraphic techniques, to the sub-mm, at which thermal emission from mm- to cm-sized
dust is seen that is thought to trace the parent planetesimals \citep[for a review, see][]{Hughes2018}.

In general the structures observed in debris disks are well described as planetesimal belts.
However, these {\it belts} range in width from narrow rings ($\Delta r/r \approx 0.1$) to broad disks
($\Delta r/r >1$).
Some of those which are broad have radial sub-structure such as gaps which may be carved by planets,
\citep[e.g.,][]{Marino2018,Marino2019,Marino2020}, reminiscent of similar structures seen in protoplanetary
disks \citep{Andrews2018}, albeit over a narrower range of radii.
The inner edges of many disks are found to be sharp, which has been inferred to be evidence for
sculpting by a planet at the inner edge \citep{Chiang2009}.
However, there are other possible interpretations, such as the radial profile being set by where planetesimals were able to form within protoplanetary disks, such as at snowlines \citep[e.g.,][]{Schoonenberg2017}.
Indeed there is emerging evidence that the radii of debris belts may correlate with snowline locations \citep{Matra2018}.

The vertical structure of disks that are not face-on can also now be resolved, with scale
heights of order a few \% being inferred and linked to stirring processes by bodies embedded
within or near to the disks \citep{Daley2019,Matra2019}.
Non-axisymmetric structures in the disks also provide evidence for planet-sized bodies, the
most unambiguous of which is the warp in the $\beta$ Pictoris disk, since that may be linked
to the known planet in the system \citep{Lagrange2010,Apai2015}.
Other asymmetric structures, such as eccentric disks, clumps and spirals \citep{Wyatt2005,Kalas2013,Dent2014}, have also been interpreted as evidence for planets.
However, more examples of systems where known planets can be linked to observed disk structures would boost confidence in what can be inferred from disk structures about underlying planetary systems.
Nevertheless, we can be confident that even small planets (e.g., as small as Neptune or even Earth) would leave an imprint in the structure of a debris disk \citep[e.g.,][]{Wyatt1999,Pearce2014,Shannon2016}, moreover one that would be readily detectable with current imaging capabilities.
Thus high resolution observations of disk structure have the potential to pinpoint the location of planets that would otherwise be extremely hard to detect using any other techniques (e.g., Neptune analogues), and also can inform about their evolutionary history \citep[e.g., providing evidence of past migration;][]{Wyatt2003}.

We are truly in an era of high resolution disk imaging, with {\it ALMA} continuing to image
new disks that can be followed up at even higher resolution.
{\it JWST} looks set to provide further high resolution images both of exo-Kuiper belts and of any (mid-planetary system) dust structures that may lie inside these.
Longer term, coronagraphic imaging with {\it WFIRST} and {\it ELT} will provide further advances.
These disk images provide our best window into outer planetary systems and the mechanisms involved in their formation and evolution.
Theoretical modelling techniques will also continue to be developed to find ways to extract more concrete evidence for the presence of planets from the images, e.g., using kinematic structures in the gas \citep{Dent2014,Teague2018}, and this field will further benefit when direct imaging capabilities allow the detection of planets toward more than a few \% of stars.

\section{Planetary bombardment} 
\label{sec:bomb}
As noted in \S \ref{sec:ss}, there remains uncertainty about the level of cometary activity in the inner few au of nearby stars, as traced by their level of exozodiacal dust.
However, it is already clear that for 20\% of stars, the level of dust in their inner regions
is orders of magnitude higher than in the Solar system.
These exozodi detections correlate with the presence of an outer belt \citep{Mennesson2014,Ertel2020},
but it is debated whether this dust is dragged inwards from the outer
belt \citep[e.g.,][]{Rigley2020} or if the material arrives in the inner regions on
larger planetesimals through scattering amongst a planetary system before being released by
sublimation or disintegration \citep[e.g.,][]{DiSisto2009,Marboeuf2016}.
It will soon be possible to test such models by using {\it JWST} (and later {\it WFIRST} and {\it ELT}) to search for dust at intermediate $\sim 10$\,au separations.
However, it is already evident that dust levels in some systems are too high to be
explained by drag processes and so planetesimal delivery is required \citep{Defrere2015,Rigley2020}.

If exozodis are replenished by planetesimals this opens up the possibility that some of these planetesimals may collide with planets in the system, an interaction with implications for conditions on the planets' surfaces.
For example, planetesimal bombardment can both deplete a pre-existing atmosphere and deliver
complex organic molecules and water, as well as volatiles that can be released to supply
a secondary atmosphere \citep{Melosh1989,Chyba1990}.
The level of planetesimal bombardment in a system depends on a number of factors, most of which
are completely unknown, such as the architecture of the outer planetary system.
While this means there is potential for constraining that planetary system from observations of
exozodiacal dust and its relation to any outer belt \citep[i.e., providing complementary information to that discussed in \S \ref{sec:arch}, e.g.,][]{Faramaz2017,Wyatt2017,Marino2018b}, there are a number of theoretical hurdles.
Notably, there is still not a complete model that links the level of zodiacal dust to its origin in comets
\citep{Nesvorny2010}.
Clearly there is scope for improved models that include all of the processes at play in debris disks, including dynamical interactions, collisions, radiation forces, gas release, gas drag, etc. \citep[e.g.,][]{Kral2013,Pearce2020}.

While the exozodiacal dust observations show that it is plausible that planetesimal bombardment is ongoing in a significant fraction of systems, it should be acknowledged that bombardment does not need to be ongoing to have (or have had) a significant effect on a planet.
High levels of warm dust are more common around stars in the first few 100\,Myr of their lives \citep[e.g.,][]{Siegler2007,Kennedy2013}, which could be indicative of an early bombardment phase as planetesimal belts are being cleared.
Indeed, the Solar system's terrestrial planets are thought to have experienced high bombardment levels early in their history \citep{Morbidelli2018}.
In general what matters for a planet is its integrated bombardment, the timing being less critical, and so it will be important to understand what these young bright warm disks tell us about typical bombardment histories.

This evidence for bombardment is opening up new lines of research, since it is now possible to consider for example how this would affect a planet's propensity for habitability.
It has been shown that there is a shoreline in planet mass versus semi-major axis space that divides planets that would be expected to grow secondary atmospheres under bombardment, and those that would be expected to have their atmospheres eroded \citep{Wyatt2020}.
This process would be competing with the effects of evaporation due
to irradiation by the star, but also with secondary outgassing as well as geological processes.
Such interactions have been considered for Solar system bodies where the bombardment history is better constrained, but it is becoming relevant to consider such processes in an exoplanetary context, given the growing body of constraints on exoplanetary atmospheres and geology.

Whereas systems with bright hot dust were discussed above in the context of external bombardment onto the planets, it has also been interpreted as evidence for ongoing planet formation processes \citep{Wyatt2016}.
Specifically, this could be the dust created in the final giant impact stage of terrestrial planet growth.
There is evidence for this in one system from the silica composition of its dust inferred from the
infrared spectrum \citep{Lisse2009}.
The rapid variability in some systems \citep{Meng2014} is also explained in this context
\citep{Jackson2014,Su2019}.
If so, observations of these extreme debris disks tell us something about the timing and
duration of the giant impact phase, and potentially about the frequency of terrestrial
planet formation \citep[e.g.,][]{Jackson2012}.
For example, for Sun-like stars this dust is seen within a few 0.1\,au of the star,
which could be because this is where planets tend to form for such stars.
This is perhaps unsurprising given the prevalence of close-in super-Earth systems seen in transit surveys,
but this could mean that Solar system-like configurations are rare.
The coming years will see more examples of such extreme disks, as well as studies of their
variability.
We can also expect advances in the models for their interpretation which need to include
a wider range of physical effects (e.g., due to the optical depth of the dust).
There are also puzzles that for now defy easy explanation, such as the rapid dispersal of one disk
\citep{Melis2012}, and the high fraction of stars with significant quantities of $\sim 1000$\,K dust \citep[e.g.,][]{Ertel2014}.
Solutions to these puzzles may come in the form of theoretical insights \citep[e.g.,][]{Rieke2016,Pearce2020} or from further observational characterisation of the phenomena.

\section{Planetesimal composition} 
\label{sec:comp}
A fundamental property of the circumstellar material is its composition.
This could provide information on where and when the planetesimals replenishing debris disk dust
formed, as well as how those bodies may have been subsequently altered by physical and geological processes.

The spectrum of the dust emission can in some cases be used to infer what it is made of.
Spectral features in the mid-IR associated with silicates are often used in this way.
However, most debris disks are too cold to emit in the mid-IR.
Such spectral features are also sensitive to the size of the dust, being present only
if the dust is smaller than the mid-IR wavelengths that characterise the feature.
Nevertheless, this has been used to show that the bright warm dust in a few systems
is silicate-rich and bears a similarity to dust compositions found in the Solar system
\citep{Lisse2012}, while in one case the abundance of silica implied formation in
a recent giant impact \citep{Lisse2009}.
{\it Herschel} opened up a new possibility for the colder disks which was to identify forsterite
features at 69\,$\mu$m, allowing the fraction of Fe and Mg to be determined \citep{deVries2012}.
However, this was only achieved for 2 disks.
Future far-IR observatories like {\it Spica} could make such measurements routine.
Scattered light observations also have the potential to probe the dust composition, e.g.,
through its colour or phase function \citep[e.g.,][]{Debes2008}, although such inferences are again prone to sensitivity to the size of the dust and are limited by how well we can model dust optical properties.
Nevertheless, those optical properties can be characterised using techniques like the discrete dipole approximation \citep[e.g.,][]{Arnold2019}, and empirical comparisons with Solar system objects are also possible \citep{Hedman2015}.
The current set of observations with instruments like {\it GPI} and {\it SPHERE} have lead to a
recent increase in the number of scattered light detections
\citep{Olofsson2020,Esposito2020}, and these will gain momentum with {\it JWST}.

The volatile content of the planetesimals is harder to probe through spectral features, though there are mid- and far-IR features due to ice that could be detected by {\it JWST} and {\it Spica} \citep{Kim2019}.
However, {\it ALMA} is providing a growing body of observations of gas in debris disks which is thought
to have a secondary origin, i.e., to have been released from volatile-rich planetesimals, rather than
being a remnant of the protoplanetary disk \citep{Dent2014,Marino2016,Marino2020b}.
Most commonly detected is CO, but there were a few detections of atomic carbon and oxygen with {\it Herschel} \citep[e.g.,][]{Cataldi2014}, and neutral carbon detections with {\it ALMA} are becoming more prevalent \citep{Higuchi2017,Cataldi2018,Kral2019,Cataldi2020}.
The paradigm within which these observations are interpreted is that of release of volatiles during
the collisional cascade process, with those molecules photodissociating relatively quickly to produce
an atomic gas disk that spreads viscously \citep{Kral2016,Kral2017}.
The relative abundances of the observed gas and dust constrain their abundances on planetesimals which
looks to be similar to Solar system comets \citep{Marino2016,Matra2017}.
This paradigm is not without its challenges, for example there remains debate about whether the gas is produced stochastically rather than in steady state \citep{Cataldi2018,Cataldi2020}.
We can expect significant observational progress on this in the coming years, first with {\it ALMA} and {\it SOFIA}, but in the future with {\it Spica}.
These observational advances will have to go hand-in-hand with improvements in the models to accurately couple the hydrodynamical evolution of the gas with its thermodynamics and chemistry.

Perhaps the most direct measurement of planetesimal composition comes from observations of
white dwarf atmospheres \citep[e.g.,][]{Farihi2016}.
These are frequently seen to be polluted with metals which have short sinking times, and so require
constant replenishment which is inferred to be from planetesimals that are tidally disrupted
before forming a disk that accretes onto the star \citep{Jura2014,Hollands2018}.
Multiple species have been observed in the atmospheres of many stars allowing the bulk composition of the pollutants to be measured, which can then be interpreted in terms of the temperature at which the planetesimal formed and the differentiation processes that it underwent since formation \citep{Bonsor2020}.
While such inferences are possible, there are still many unknowns; e.g., there is no
direct evidence for the location of the parent planetesimal belt \citep{Farihi2014}, the mechanism which
caused the planetesimal to arrive at the star is not known, as are the details of the tidal
disruption and disk accretion process \citep{Veras2016}. 
With large numbers of white dwarfs being discovered in the era of large all-sky surveys like {\it Gaia}, {\it SDSS}, {\it ZTF} and {\it LSST} \citep{GentileFusillo2019}, and follow-up spectroscopy being performed to characterise any pollution, we can anticipate new observational results in the coming years, as well as significant theoretical progress on the various steps needed for the interpretation.

\section{Protoplanetary disk dispersal} 
\label{sec:disp}
The recent far-IR surveys of {\it Spitzer} and {\it Herschel} have reinforced the paradigm in which stars are born with a planetesimal belt that subsequently erodes through collisional processing that passes mass from large planetesimals down to dust that is eventually expelled by radiation forces \citep{Wyatt2007,Marino2020}.
This model fits the statistics for the fraction of stars seen to have detectable emission as a function of age in different wavebands.
However, as already noted this necessarily cannot explain the 80\% of stars without detectable disks.
Further analysis of the current detections is also leading to new insights, such as a correlation
of planetesimal belt radius with stellar mass \citep{Matra2018}.
The model also avoids the question of how and when the planetesimal belts formed, rather presupposing
that these are present, with a collisional cascade already set up, as soon as the protoplanetary disk
disperses.
However, with the discovery of gas in debris disks, and the knowledge that dust in protoplanetary disks
is not primordial but subject to a continual cycle of growth and destruction, it is becoming clear
that the distinction between the two types of disk is not well defined \citep{Wyatt2015}.
This is particularly noticeable when considering the youngest systems, $<20$\,Myr or so, which means
that observations of such young stars can inform on the processes of protoplanetary disk dispersal
and of the birth of a debris disk.

Studies of young stars generally involve first identifying associations of stars that appear
co-eval and with similar distances and space motions.
However, the relevant kinematic information has until now not been available
for the nearest star forming regions which lie at $\sim 150$\,pc.
While it has nevertheless been relatively unambiguous to identify protoplanetary disks in this way
(i.e., the class II stars), this is harder for the young stars without such an obviously bright disk
(i.e., the class III stars), with background stars often mistakenly falling in this category \citep{Manara2018}.
Gaia is changing that, and the coming years will see improved membership of the different nearby
associations from which to study their disk populations \citep{Canovas2019,Luhman2020}.
At the same time, {\it ALMA} observations of such regions have until now been relatively shallow, i.e., only sensitive to protoplanetary disks \citep[e.g.,][]{Ansdell2016,Williams2019}.
However, deeper {\it ALMA} observations can probe for dust emission at levels of the brightest debris disks in the nearest star forming regions, and can also search for remnant primordial gas that may have yet to disperse after the dust has been depleted \citep[e.g.,][]{Owen2019}.
Thus it is timely now to perform a census of the youngest debris disks as well as to search for protoplanetary disks in the final stages of dissipation \citep[e.g.,][]{Lovell2021}.
Far-IR surveys are still more sensitive than those in the sub-mm, and for those we
will have to wait for {\it Spica} which would be able to set limits on the presence of dust that are comparable with those for nearby stars (except where imaging has been able to probe fainter dust, see \S \ref{sec:ss}).

Significant advances have also been made in our understanding of protoplanetary disks, both in their observational characterisation and in theoretical advances in modelling the processes within them \citep[for a recent review, see][]{Andrews2020}.
These will continue along with the advances in debris disks described in this chapter, and together it should be possible to infer the sequence of events that lead to the depletion of a protoplanetary disk.
Observations of the gas will be particularly telling, since this component dominates the mass of a protoplanetary disk, and its dynamical influence is responsible for much of the dust morphology in such disks.
Observations of the dispersal of that component in a disk wind would constrain that process \citep{Haworth2020},
as well as its impact on dust and planetesimals in the disk \citep{Carrera2017, Owen2019}, and planet formation more generally.

\section{Opportunities and challenges} 
Debris disks represent an opportunity, by providing unique information about exoplanetary systems.
Observationally one challenge is to detect debris disks as faint as those in the Solar
System, particularly given how relatively bright the host star is.
Theoretically the challenge is to extract reliable information in the face of so many unknowns about
the architecture, history and detailed physics of the underlying planetary system.

\bibliographystyle{plain}
\bibliography{refs}

\begin{thebibliography}{100}

\bibitem{Alvarez1980}
Luis~W. {Alvarez}, Walter {Alvarez}, Frank {Asaro}, and Helen~V. {Michel}.
\newblock {Extraterrestrial Cause for the Cretaceous-Tertiary Extinction}.
\newblock {\em Science}, 208:1095, 1980.

\bibitem{Andrews2020}
Sean~M. Andrews.
\newblock Observations of protoplanetary disk structures.
\newblock {\em ARA\&A}, 58:483, 2020.

\bibitem{Andrews2018}
Sean~M. {Andrews}, Jane {Huang}, Laura~M. {P{\'e}rez}, Andrea {Isella},
  Cornelis~P. {Dullemond}, Nicol{\'a}s~T. {Kurtovic}, Viviana~V. {Guzm{\'a}n},
  John~M. {Carpenter}, David~J. {Wilner}, Shangjia {Zhang}, Zhaohuan {Zhu},
  Tilman {Birnstiel}, Xue-Ning {Bai}, Myriam {Benisty}, A.~Meredith {Hughes},
  Karin~I. {{\"O}berg}, and Luca {Ricci}.
\newblock {The Disk Substructures at High Angular Resolution Project (DSHARP).
  I. Motivation, Sample, Calibration, and Overview}.
\newblock {\em ApJL}, 869:L41, 2018.

\bibitem{Ansdell2016}
M.~{Ansdell}, J.~P. {Williams}, N.~{van der Marel}, J.~M. {Carpenter},
  G.~{Guidi}, M.~{Hogerheijde}, G.~S. {Mathews}, C.~F. {Manara}, A.~{Miotello},
  A.~{Natta}, I.~{Oliveira}, M.~{Tazzari}, L.~{Testi}, E.~F. {van Dishoeck},
  and S.~E. {van Terwisga}.
\newblock {ALMA Survey of Lupus Protoplanetary Disks. I. Dust and Gas Masses}.
\newblock {\em ApJ}, 828:46, 2016.

\bibitem{Apai2015}
D{\'a}niel {Apai}, Glenn {Schneider}, Carol~A. {Grady}, Mark~C. {Wyatt},
  Anne-Marie {Lagrange}, Marc~J. {Kuchner}, Christopher~J. {Stark}, and
  Stephen~H. {Lubow}.
\newblock {The Inner Disk Structure, Disk-Planet Interactions, and Temporal
  Evolution in the {\ensuremath{\beta}} Pictoris System: A Two-epoch HST/STIS
  Coronagraphic Study}.
\newblock {\em ApJ}, 800:136, 2015.

\bibitem{Arnold2019}
Jessica~A. {Arnold}, Alycia~J. {Weinberger}, Gorden {Videen}, and Evgenij~S.
  {Zubko}.
\newblock {The Effect of Dust Composition and Shape on Radiation-pressure
  Forces and Blowout Sizes of Particles in Debris Disks}.
\newblock {\em AJ}, 157:157, 2019.

\bibitem{Aumann1984}
H.~H. {Aumann}, F.~C. {Gillett}, C.~A. {Beichman}, T.~{de Jong}, J.~R. {Houck},
  F.~J. {Low}, G.~{Neugebauer}, R.~G. {Walker}, and P.~R. {Wesselius}.
\newblock {Discovery of a shell around alpha Lyrae}.
\newblock {\em ApJL}, 278:L23, 1984.

\bibitem{Bonsor2020}
Amy {Bonsor}, Philip~J. {Carter}, Mark {Hollands}, Boris~T. {G{\"a}nsicke},
  Zo{\"e} {Leinhardt}, and John H.~D. {Harrison}.
\newblock {Are exoplanetesimals differentiated?}
\newblock {\em MNRAS}, 492:2683, 2020.

\bibitem{Booth2009}
Mark {Booth}, Mark~C. {Wyatt}, Alessandro {Morbidelli}, Amaya
  {Moro-Mart{\'\i}n}, and Harold~F. {Levison}.
\newblock {The history of the Solar system's debris disc: observable properties
  of the Kuiper belt}.
\newblock {\em MNRAS}, 399:385, 2009.

\bibitem{Canovas2019}
H.~{C{\'a}novas}, C.~{Cantero}, L.~{Cieza}, A.~{Bombrun}, U.~{Lammers},
  B.~{Mer{\'\i}n}, A.~{Mora}, {\'A}.~{Ribas}, and
  D.~{Ru{\'\i}z-Rodr{\'\i}guez}.
\newblock {Census of {\ensuremath{\rho}} Ophiuchi candidate members from Gaia
  Data Release 2}.
\newblock {\em A\&A}, 626:A80, 2019.

\bibitem{Carrera2017}
Daniel {Carrera}, Uma {Gorti}, Anders {Johansen}, and Melvyn~B. {Davies}.
\newblock {Planetesimal Formation by the Streaming Instability in a
  Photoevaporating Disk}.
\newblock {\em ApJ}, 839:16, 2017.

\bibitem{Cataldi2014}
G.~{Cataldi}, A.~{Brandeker}, G.~{Olofsson}, B.~{Larsson}, R.~{Liseau},
  J.~{Blommaert}, M.~{Fridlund}, R.~{Ivison}, E.~{Pantin}, B.~{Sibthorpe},
  B.~{Vandenbussche}, and Y.~{Wu}.
\newblock {Herschel/HIFI observations of ionised carbon in the
  {\ensuremath{\beta}} Pictoris debris disk}.
\newblock {\em A\&A}, 563:A66, 2014.

\bibitem{Cataldi2018}
Gianni {Cataldi}, Alexis {Brandeker}, Yanqin {Wu}, Christine {Chen}, William
  {Dent}, Bernard~L. {de Vries}, Inga {Kamp}, Ren{\'e} {Liseau}, G{\"o}ran
  {Olofsson}, Eric {Pantin}, and Aki {Roberge}.
\newblock {ALMA Resolves C I Emission from the {\ensuremath{\beta}} Pictoris
  Debris Disk}.
\newblock {\em ApJ}, 861:72, 2018.

\bibitem{Cataldi2020}
Gianni {Cataldi}, Yanqin {Wu}, Alexis {Brandeker}, Nagayoshi {Ohashi}, Attila
  {Mo{\'o}r}, G{\"o}ran {Olofsson}, P{\'e}ter {{\'A}brah{\'a}m}, Ruben
  {Asensio-Torres}, Maria {Cavallius}, William R.~F. {Dent}, Carol {Grady},
  Thomas {Henning}, Aya~E. {Higuchi}, A.~Meredith {Hughes}, Markus {Janson},
  Inga {Kamp}, {\'A}gnes {K{\'o}sp{\'a}l}, Seth {Redfield}, Aki {Roberge},
  Alycia {Weinberger}, and Barry {Welsh}.
\newblock {The Surprisingly Low Carbon Mass in the Debris Disk around HD
  32297}.
\newblock {\em ApJ}, 892:99, 2020.

\bibitem{Chiang2009}
E.~{Chiang}, E.~{Kite}, P.~{Kalas}, J.~R. {Graham}, and M.~{Clampin}.
\newblock {Fomalhaut's Debris Disk and Planet: Constraining the Mass of
  Fomalhaut b from disk Morphology}.
\newblock {\em ApJ}, 693:734, 2009.

\bibitem{Chyba1990}
C.~F. {Chyba}.
\newblock {Impact delivery and erosion of planetary oceans in the early inner
  Solar System}.
\newblock {\em Nature}, 343:129, 1990.

\bibitem{Daley2019}
Cail {Daley}, A.~Meredith {Hughes}, Evan~S. {Carter}, Kevin {Flaherty}, Zachary
  {Lambros}, Margaret {Pan}, Hilke {Schlichting}, Eugene {Chiang}, Mark
  {Wyatt}, David {Wilner}, Sean {Andrews}, and John {Carpenter}.
\newblock {The Mass of Stirring Bodies in the AU Mic Debris Disk Inferred from
  Resolved Vertical Structure}.
\newblock {\em ApJ}, 875:87, 2019.

\bibitem{deVries2012}
B.~L. {de Vries}, B.~{Acke}, J.~A.~D.~L. {Blommaert}, C.~{Waelkens},
  L.~B.~F.~M. {Waters}, B.~{Vandenbussche}, M.~{Min}, G.~{Olofsson},
  C.~{Dominik}, L.~{Decin}, M.~J. {Barlow}, A.~{Brandeker}, J.~{di Francesco},
  A.~M. {Glauser}, J.~{Greaves}, P.~M. {Harvey}, W.~S. {Holland }, R.~J.
  {Ivison}, R.~{Liseau}, E.~E. {Pantin}, G.~L. {Pilbratt}, P.~{Royer}, and
  B.~{Sibthorpe}.
\newblock {Comet-like mineralogy of olivine crystals in an extrasolar
  proto-Kuiper belt}.
\newblock {\em Nature}, 490:74, 2012.

\bibitem{Debes2008}
John~H. {Debes}, Alycia~J. {Weinberger}, and Glenn {Schneider}.
\newblock {Complex Organic Materials in the Circumstellar Disk of HR 4796A}.
\newblock {\em ApJL}, 673:L191, 2008.

\bibitem{Defrere2015}
D.~{Defr{\`e}re}, P.~M. {Hinz}, A.~J. {Skemer}, G.~M. {Kennedy}, V.~P.
  {Bailey}, W.~F. {Hoffmann}, B.~{Mennesson}, R.~{Millan-Gabet}, W.~C.
  {Danchi}, O.~{Absil}, P.~{Arbo}, C.~{Beichman}, G.~{Brusa}, G.~{Bryden},
  E.~C. {Downey}, O.~{Durney}, S.~{Esposito}, A.~{Gaspar}, P.~{Grenz},
  C.~{Haniff}, J.~M. {Hill}, J.~{Lebreton}, J.~M. {Leisenring}, J.~R. {Males},
  L.~{Marion}, T.~J. {McMahon}, M.~{Montoya}, K.~M. {Morzinski}, E.~{Pinna},
  A.~{Puglisi}, G.~{Rieke}, A.~{Roberge}, E.~{Serabyn}, R.~{Sosa},
  K.~{Stapeldfeldt}, K.~{Su}, V.~{Vaitheeswaran}, A.~{Vaz}, A.~J. {Weinberger},
  and M.~C. {Wyatt}.
\newblock {First-light LBT Nulling Interferometric Observations: Warm
  Exozodiacal Dust Resolved within a Few AU of {\ensuremath{\eta}} Crv}.
\newblock {\em ApJ}, 799:42, 2015.

\bibitem{Dent2014}
W.~R.~F. {Dent}, M.~C. {Wyatt}, A.~{Roberge}, J.~C. {Augereau}, S.~{Casassus},
  S.~{Corder}, J.~S. {Greaves}, I.~{de Gregorio-Monsalvo}, A.~{Hales}, A.~P.
  {Jackson}, A.~Meredith {Hughes}, A.~M. {Lagrange}, B.~{Matthews}, and
  D.~{Wilner}.
\newblock {Molecular Gas Clumps from the Destruction of Icy Bodies in the
  {\ensuremath{\beta}} Pictoris Debris Disk}.
\newblock {\em Science}, 343:1490, 2014.

\bibitem{DiSisto2009}
Romina~P. {Di Sisto}, Julio~A. {Fern{\'a}ndez}, and Adri{\'a}n {Brunini}.
\newblock {On the population, physical decay and orbital distribution of
  Jupiter family comets: Numerical simulations}.
\newblock {\em Icarus}, 203:140, 2009.

\bibitem{Eiroa2013}
C.~{Eiroa}, J.~P. {Marshall}, A.~{Mora}, B.~{Montesinos}, O.~{Absil}, J.~Ch.
  {Augereau}, A.~{Bayo}, G.~{Bryden}, W.~{Danchi}, C.~{del Burgo}, S.~{Ertel},
  M.~{Fridlund}, A.~M. {Heras}, A.~V. {Krivov}, R.~{Launhardt}, R.~{Liseau},
  T.~{L{\"o}hne}, J.~{Maldonado}, G.~L. {Pilbratt}, A.~{Roberge}, J.~{Rodmann},
  J.~{Sanz-Forcada}, E.~{Solano}, K.~{Stapelfeldt}, P.~{Th{\'e}bault},
  S.~{Wolf}, D.~{Ardila}, M.~{Ar{\'e}valo}, C.~{Beichmann}, V.~{Faramaz}, B.~M.
  {Gonz{\'a}lez-Garc{\'\i}a}, R.~{Guti{\'e}rrez}, J.~{Lebreton},
  R.~{Mart{\'\i}nez-Arn{\'a}iz}, G.~{Meeus}, D.~{Montes}, G.~{Olofsson},
  K.~Y.~L. {Su}, G.~J. {White}, D.~{Barrado}, M.~{Fukagawa}, E.~{Gr{\"u}n},
  I.~{Kamp}, R.~{Lorente}, A.~{Morbidelli}, S.~{M{\"u}ller}, H.~{Mutschke},
  T.~{Nakagawa}, I.~{Ribas}, and H.~{Walker}.
\newblock {DUst around NEarby Stars. The survey observational results}.
\newblock {\em A\&A}, 555:A11, 2013.

\bibitem{Ertel2014}
S.~{Ertel}, O.~{Absil}, D.~{Defr{\`e}re}, J.~B. {Le Bouquin}, J.~C. {Augereau},
  L.~{Marion}, N.~{Blind}, A.~{Bonsor}, G.~{Bryden}, J.~{Lebreton}, and
  J.~{Milli}.
\newblock {A near-infrared interferometric survey of debris-disk stars. IV. An
  unbiased sample of 92 southern stars observed in H band with VLTI/PIONIER}.
\newblock {\em A\&A}, 570:A128, 2014.

\bibitem{Ertel2020}
S.~{Ertel}, D.~{Defr{\`e}re}, P.~{Hinz}, B.~{Mennesson}, G.~M. {Kennedy}, W.~C.
  {Danchi}, C.~{Gelino}, J.~M. {Hill}, W.~F. {Hoffmann}, J.~{Mazoyer},
  G.~{Rieke}, A.~{Shannon}, K.~{Stapelfeldt}, E.~{Spalding}, J.~M. {Stone},
  A.~{Vaz}, A.~J. {Weinberger}, P.~{Willems}, O.~{Absil}, P.~{Arbo}, V.~P.
  {Bailey}, C.~{Beichman}, G.~{Bryden}, E.~C. {Downey}, O.~{Durney},
  S.~{Esposito}, A.~{Gaspar}, P.~{Grenz}, C.~A. {Haniff}, J.~M. {Leisenring},
  L.~{Marion}, T.~J. {McMahon}, R.~{Millan-Gabet}, M.~{Montoya}, K.~M.
  {Morzinski}, S.~{Perera}, E.~{Pinna}, J.~U. {Pott}, J.~{Power}, A.~{Puglisi},
  A.~{Roberge}, E.~{Serabyn}, A.~J. {Skemer}, K.~Y.~L. {Su},
  V.~{Vaitheeswaran}, and M.~C. {Wyatt}.
\newblock {The HOSTS Survey for Exozodiacal Dust: Observational Results from
  the Complete Survey}.
\newblock {\em AJ}, 159:177, 2020.

\bibitem{Esposito2020}
Thomas~M. {Esposito}, Paul {Kalas}, Michael~P. {Fitzgerald}, Maxwell~A.
  {Millar-Blanchaer}, Gaspard {Duch{\^e}ne}, Jennifer {Patience}, Justin {Hom},
  Marshall~D. {Perrin}, Robert~J. {De Rosa}, Eugene {Chiang}, Ian {Czekala},
  Bruce {Macintosh}, James~R. {Graham}, Megan {Ansdell}, Pauline {Arriaga},
  Sebastian {Bruzzone}, Joanna {Bulger}, Christine~H. {Chen}, Tara {Cotten},
  Ruobing {Dong}, Zachary~H. {Draper}, Katherine~B. {Follette}, Li-Wei {Hung},
  Ronald {Lopez}, Brenda~C. {Matthews}, Johan {Mazoyer}, Stan {Metchev}, Julien
  {Rameau}, Bin {Ren}, Malena {Rice}, Inseok {Song}, Kevin {Stahl}, Jason
  {Wang}, Schuyler {Wolff}, Ben {Zuckerman}, S.~Mark {Ammons}, Vanessa~P.
  {Bailey}, Travis {Barman}, Jeffrey {Chilcote}, Rene {Doyon}, Benjamin~L.
  {Gerard}, Stephen~J. {Goodsell}, Alexandra~Z. {Greenbaum}, Pascale {Hibon},
  Sasha {Hinkley}, Patrick {Ingraham}, Quinn {Konopacky}, J{\'e}r{\^o}me
  {Maire}, Franck {Marchis}, Mark~S. {Marley}, Christian {Marois}, Eric~L.
  {Nielsen}, Rebecca {Oppenheimer}, David {Palmer}, Lisa {Poyneer}, Laurent
  {Pueyo}, Abhijith {Rajan}, Fredrik~T. {Rantakyr{\"o}}, Jean-Baptiste
  {Ruffio}, Dmitry {Savransky}, Adam~C. {Schneider}, Anand {Sivaramakrishnan},
  R{\'e}mi {Soummer}, Sandrine {Thomas}, and Kimberly {Ward-Duong}.
\newblock {Debris Disk Results from the Gemini Planet Imager Exoplanet Survey's
  Polarimetric Imaging Campaign}.
\newblock {\em AJ}, 160:24, 2020.

\bibitem{Faramaz2017}
V.~{Faramaz}, S.~{Ertel}, M.~{Booth}, J.~{Cuadra}, and C.~{Simmonds}.
\newblock {Inner mean-motion resonances with eccentric planets: a possible
  origin for exozodiacal dust clouds}.
\newblock {\em MNRAS}, 465:2352, 2017.

\bibitem{Farihi2016}
J.~{Farihi}.
\newblock {Circumstellar debris and pollution at white dwarf stars}.
\newblock {\em NewAR}, 71:9, 2016.

\bibitem{Farihi2014}
J.~{Farihi}, M.~C. {Wyatt}, J.~S. {Greaves}, A.~{Bonsor}, B.~{Sibthorpe}, and
  O.~{Pani{\'c}}.
\newblock {ALMA and Herschel observations of the prototype dusty and polluted
  white dwarf G29-38}.
\newblock {\em MNRAS}, 444:1821, 2014.

\bibitem{Gaspar2013}
Andr{\'a}s {G{\'a}sp{\'a}r}, George~H. {Rieke}, and Zolt{\'a}n {Balog}.
\newblock {The Collisional Evolution of Debris Disks}.
\newblock {\em ApJ}, 768:25, 2013.

\bibitem{GentileFusillo2019}
Nicola~Pietro {Gentile Fusillo}, Pier-Emmanuel {Tremblay}, Boris~T.
  {G{\"a}nsicke}, Christopher~J. {Manser}, Tim {Cunningham}, Elena
  {Cukanovaite}, Mark {Hollands}, Thomas {Marsh}, Roberto {Raddi}, Stefan
  {Jordan}, Silvia {Toonen}, Stephan {Geier}, Martin {Barstow}, and Jeffrey~D.
  {Cummings}.
\newblock {A Gaia Data Release 2 catalogue of white dwarfs and a comparison
  with SDSS}.
\newblock {\em MNRAS}, 482:4570, 2019.

\bibitem{Greaves1998}
J.~S. {Greaves}, W.~S. {Holland}, G.~{Moriarty-Schieven}, T.~{Jenness},
  W.~R.~F. {Dent}, B.~{Zuckerman}, C.~{McCarthy}, R.~A. {Webb}, H.~M. {Butner},
  W.~K. {Gear}, and H.~J. {Walker}.
\newblock {A Dust Ring around {\ensuremath{\in}} Eridani: Analog to the Young
  Solar System}.
\newblock {\em ApJL}, 506:L133, 1998.

\bibitem{Haworth2020}
Thomas~J. {Haworth} and James~E. {Owen}.
\newblock {The observational anatomy of externally photoevaporating
  planet-forming discs - I. Atomic carbon}.
\newblock {\em MNRAS}, 492:5030, 2020.

\bibitem{Hedman2015}
Matthew~M. {Hedman} and Christopher~C. {Stark}.
\newblock {Saturn's G and D Rings Provide Nearly Complete Measured Scattering
  Phase Functions of Nearby Debris Disks}.
\newblock {\em ApJ}, 811:67, 2015.

\bibitem{Higuchi2017}
Aya~E. {Higuchi}, Aki {Sato}, Takashi {Tsukagoshi}, Nami {Sakai}, Kazunari
  {Iwasaki}, Munetake {Momose}, Hiroshi {Kobayashi}, Daisuke {Ishihara}, Sakae
  {Watanabe}, Hidehiro {Kaneda}, and Satoshi {Yamamoto}.
\newblock {Detection of Submillimeter-wave [C I] Emission in Gaseous Debris
  Disks of 49 Ceti and {\ensuremath{\beta}} Pictoris}.
\newblock {\em ApJL}, 839:L14, 2017.

\bibitem{Holland1998}
Wayne~S. {Holland}, Jane~S. {Greaves}, B.~{Zuckerman}, R.~A. {Webb}, Chris
  {McCarthy}, Iain~M. {Coulson}, D.~M. {Walther}, William R.~F. {Dent},
  Walter~K. {Gear}, and Ian {Robson}.
\newblock {Submillimetre images of dusty debris around nearby stars}.
\newblock {\em Nature}, 392:788, 1998.

\bibitem{Hollands2018}
M.~A. {Hollands}, B.~T. {G{\"a}nsicke}, and D.~{Koester}.
\newblock {Cool DZ white dwarfs II: compositions and evolution of old remnant
  planetary systems}.
\newblock {\em MNRAS}, 477:93, 2018.

\bibitem{Hughes2018}
A.~Meredith {Hughes}, Gaspard {Duch{\^e}ne}, and Brenda~C. {Matthews}.
\newblock {Debris Disks: Structure, Composition, and Variability}.
\newblock {\em ARA\&A}, 56:541, 2018.

\bibitem{Jackson2012}
Alan~P. {Jackson} and Mark~C. {Wyatt}.
\newblock {Debris from terrestrial planet formation: the Moon-forming
  collision}.
\newblock {\em MNRAS}, 425:657, 2012.

\bibitem{Jackson2014}
Alan~P. {Jackson}, Mark~C. {Wyatt}, Amy {Bonsor}, and Dimitri {Veras}.
\newblock {Debris froms giant impacts between planetary embryos at large
  orbital radii}.
\newblock {\em MNRAS}, 440:3757, 2014.

\bibitem{Jayawardhana1998}
Ray {Jayawardhana}, Scott {Fisher}, Lee {Hartmann}, Charles {Telesco}, Robert
  {Pi{\~n}a}, and Giovanni {Fazio}.
\newblock {A Dust Disk Surrounding the Young A Star HR 4796A}.
\newblock {\em ApJL}, 503:L79, 1998.

\bibitem{Jura2014}
M.~{Jura} and E.~D. {Young}.
\newblock {Extrasolar Cosmochemistry}.
\newblock {\em AREPS}, 42:45, 2014.

\bibitem{Kalas2013}
Paul {Kalas}, James~R. {Graham}, Michael~P. {Fitzgerald}, and Mark {Clampin}.
\newblock {STIS Coronagraphic Imaging of Fomalhaut: Main Belt Structure and the
  Orbit of Fomalhaut b}.
\newblock {\em ApJ}, 775:56, 2013.

\bibitem{Kennedy2013}
G.~M. {Kennedy} and M.~C. {Wyatt}.
\newblock {The bright end of the exo-Zodi luminosity function: disc evolution
  and implications for exo-Earth detectability}.
\newblock {\em MNRAS}, 433:2334, 2013.

\bibitem{Kim2019}
M.~{Kim}, S.~{Wolf}, A.~{Potapov}, H.~{Mutschke}, and C.~{J{\"a}ger}.
\newblock {Constraining the detectability of water ice in debris disks}.
\newblock {\em A\&A}, 629:A141, 2019.

\bibitem{Kral2013}
Q.~{Kral}, P.~{Th{\'e}bault}, and S.~{Charnoz}.
\newblock {LIDT-DD: A new self-consistent debris disc model that includes
  radiation pressure and couples dynamical and collisional evolution}.
\newblock {\em A\&A}, 558:A121, 2013.

\bibitem{Kral2016}
Q.~{Kral}, M.~{Wyatt}, R.~F. {Carswell}, J.~E. {Pringle}, L.~{Matr{\`a}}, and
  A.~{Juh{\'a}sz}.
\newblock {A self-consistent model for the evolution of the gas produced in the
  debris disc of {\ensuremath{\beta}} Pictoris}.
\newblock {\em MNRAS}, 461:845, 2016.

\bibitem{Kral2019}
Quentin {Kral}, Sebastian {Marino}, Mark~C. {Wyatt}, Mihkel {Kama}, and Luca
  {Matr{\`a}}.
\newblock {Imaging [CI] around HD 131835: reinterpreting young debris discs
  with protoplanetary disc levels of CO gas as shielded secondary discs}.
\newblock {\em MNRAS}, 489:3670, 2019.

\bibitem{Kral2017}
Quentin {Kral}, Luca {Matr{\`a}}, Mark~C. {Wyatt}, and Grant~M. {Kennedy}.
\newblock {Predictions for the secondary CO, C and O gas content of debris
  discs from the destruction of volatile-rich planetesimals}.
\newblock {\em MNRAS}, 469:521, 2017.

\bibitem{Lagrange2010}
A.~M. {Lagrange}, M.~{Bonnefoy}, G.~{Chauvin}, D.~{Apai}, D.~{Ehrenreich},
  A.~{Boccaletti}, D.~{Gratadour}, D.~{Rouan}, D.~{Mouillet}, S.~{Lacour}, and
  M.~{Kasper}.
\newblock {A Giant Planet Imaged in the Disk of the Young Star
  {\ensuremath{\beta}} Pictoris}.
\newblock {\em Science}, 329:57, 2010.

\bibitem{Liseau2013}
R.~{Liseau}, B.~{Montesinos}, G.~{Olofsson}, G.~{Bryden}, J.~P. {Marshall},
  D.~{Ardila}, A.~{Bayo Aran}, W.~C. {Danchi}, C.~{del Burgo}, C.~{Eiroa},
  S.~{Ertel}, M.~C.~W. {Fridlund}, A.~V. {Krivov}, G.~L. {Pilbratt},
  A.~{Roberge}, P.~{Th{\'e}bault}, J.~{Wiegert}, and G.~J. {White}.
\newblock {{\ensuremath{\alpha}} Centauri A in the far infrared. First
  measurement of the temperature minimum of a star other than the Sun}.
\newblock {\em A\&A}, 549:L7, 2013.

\bibitem{Lisse2009}
C.~M. {Lisse}, C.~H. {Chen}, M.~C. {Wyatt}, A.~{Morlok}, I.~{Song},
  G.~{Bryden}, and P.~{Sheehan}.
\newblock {Abundant Circumstellar Silica Dust and SiO Gas Created by a Giant
  Hypervelocity Collision in the \raisebox{-0.5ex}\textasciitilde12 Myr
  HD172555 System}.
\newblock {\em ApJ}, 701:2019, 2009.

\bibitem{Lisse2012}
C.~M. {Lisse}, M.~C. {Wyatt}, C.~H. {Chen}, A.~{Morlok}, D.~M. {Watson},
  P.~{Manoj}, P.~{Sheehan}, T.~M. {Currie}, P.~{Thebault}, and M.~L. {Sitko}.
\newblock {Spitzer Evidence for a Late-heavy Bombardment and the Formation of
  Ureilites in {\ensuremath{\eta}} Corvi at \raisebox{-0.5ex}\textasciitilde1
  Gyr}.
\newblock {\em ApJ}, 747:93, 2012.

\bibitem{Lohne2008}
Torsten {L{\"o}hne}, Alexander~V. {Krivov}, and Jens {Rodmann}.
\newblock {Long-Term Collisional Evolution of Debris Disks}.
\newblock {\em ApJ}, 673:1123, 2008.

\bibitem{Lovell2021}
J.~B. {Lovell}, M.~C. {Wyatt}, M.~{Ansdell}, M.~{Kama}, G.~M. {Kennedy}, C.~F.
  {Manara}, S.~{Marino}, L.~{Matr{\`a}}, G.~{Rosotti}, M.~{Tazzari},
  L.~{Testi}, and J.~P. {Williams}.
\newblock {ALMA survey of Lupus class III stars: Early planetesimal belt
  formation and rapid disc dispersal}.
\newblock {\em MNRAS}, 500:4878, 2021.

\bibitem{Luhman2020}
K.~L. {Luhman}.
\newblock {A Gaia Survey for Young Stars Associated with the Lupus Clouds}.
\newblock {\em AJ}, 160:186, 2020.

\bibitem{Manara2018}
C.~F. {Manara}, T.~{Prusti}, F.~{Comeron}, R.~{Mor}, J.~M. {Alcal{\'a}},
  T.~{Antoja}, S.~{Facchini}, D.~{Fedele}, A.~{Frasca}, T.~{Jerabkova},
  G.~{Rosotti}, L.~{Spezzi}, and L.~{Spina}.
\newblock {Gaia DR2 view of the Lupus V-VI clouds: The candidate diskless young
  stellar objects are mainly background contaminants}.
\newblock {\em A\&A}, 615:L1, 2018.

\bibitem{Marboeuf2016}
U.~{Marboeuf}, A.~{Bonsor}, and J.~C. {Augereau}.
\newblock {Extrasolar comets: The origin of dust in exozodiacal disks?}
\newblock {\em P\&SS}, 133:47, 2016.

\bibitem{Marino2018}
S.~{Marino}, J.~{Carpenter}, M.~C. {Wyatt}, M.~{Booth}, S.~{Casassus},
  V.~{Faramaz}, V.~{Guzman}, A.~M. {Hughes}, A.~{Isella}, G.~M. {Kennedy},
  L.~{Matr{\`a}}, L.~{Ricci}, and S.~{Corder}.
\newblock {A gap in the planetesimal disc around HD 107146 and asymmetric warm
  dust emission revealed by ALMA}.
\newblock {\em MNRAS}, 479:5423, 2018.

\bibitem{Marino2020b}
S.~{Marino}, M.~{Flock}, Th~{Henning}, Q.~{Kral}, L.~{Matr{\`a}}, and M.~C.
  {Wyatt}.
\newblock {Population synthesis of exocometary gas around A stars}.
\newblock {\em MNRAS}, 492:4409, 2020.

\bibitem{Marino2016}
S.~{Marino}, L.~{Matr{\`a}}, C.~{Stark}, M.~C. {Wyatt}, S.~{Casassus},
  G.~{Kennedy}, D.~{Rodriguez}, B.~{Zuckerman}, S.~{Perez}, W.~R.~F. {Dent},
  M.~{Kuchner}, A.~M. {Hughes}, G.~{Schneider}, A.~{Steele}, A.~{Roberge},
  J.~{Donaldson}, and E.~{Nesvold}.
\newblock {Exocometary gas in the HD 181327 debris ring}.
\newblock {\em MNRAS}, 460:2933, 2016.

\bibitem{Marino2019}
S.~{Marino}, B.~{Yelverton}, M.~{Booth}, V.~{Faramaz}, G.~M. {Kennedy},
  L.~{Matr{\`a}}, and M.~C. {Wyatt}.
\newblock {A gap in HD 92945's broad planetesimal disc revealed by ALMA}.
\newblock {\em MNRAS}, 484:1257, 2019.

\bibitem{Marino2020}
S.~{Marino}, A.~{Zurlo}, V.~{Faramaz}, J.~{Milli}, Th~{Henning}, G.~M.
  {Kennedy}, L.~{Matr{\`a}}, S.~{P{\'e}rez}, P.~{Delorme}, L.~A. {Cieza}, and
  A.~M. {Hughes}.
\newblock {Insights into the planetary dynamics of HD 206893 with ALMA}.
\newblock {\em MNRAS}, 498:1319, 2020.

\bibitem{Marino2018b}
Sebastian {Marino}, Amy {Bonsor}, Mark~C. {Wyatt}, and Quentin {Kral}.
\newblock {Scattering of exocomets by a planet chain: exozodi levels and the
  delivery of cometary material to inner planets}.
\newblock {\em MNRAS}, 479:1651, 2018.

\bibitem{Matra2017}
L.~{Matr{\`a}}, W.~R.~F. {Dent}, M.~C. {Wyatt}, Q.~{Kral}, D.~J. {Wilner},
  O.~{Pani{\'c}}, A.~M. {Hughes}, I.~{de Gregorio-Monsalvo}, A.~{Hales}, J.~C.
  {Augereau}, J.~{Greaves}, and A.~{Roberge}.
\newblock {Exocometary gas structure, origin and physical properties around
  {\ensuremath{\beta}} Pictoris through ALMA CO multitransition observations}.
\newblock {\em MNRAS}, 464:1415, 2017.

\bibitem{Matra2018}
L.~{Matr{\`a}}, S.~{Marino}, G.~M. {Kennedy}, M.~C. {Wyatt}, K.~I. {{\"O}berg},
  and D.~J. {Wilner}.
\newblock {An Empirical Planetesimal Belt Radius-Stellar Luminosity Relation}.
\newblock {\em ApJ}, 859:72, 2018.

\bibitem{Matra2019}
L.~{Matr{\`a}}, M.~C. {Wyatt}, D.~J. {Wilner}, W.~R.~F. {Dent}, S.~{Marino},
  G.~M. {Kennedy}, and J.~{Milli}.
\newblock {Kuiper Belt-like Hot and Cold Populations of Planetesimal
  Inclinations in the {\ensuremath{\beta}} Pictoris Belt Revealed by ALMA}.
\newblock {\em AJ}, 157:135, 2019.

\bibitem{Melis2012}
Carl {Melis}, B.~{Zuckerman}, Joseph~H. {Rhee}, Inseok {Song}, Simon~J.
  {Murphy}, and Michael~S. {Bessell}.
\newblock {Rapid disappearance of a warm, dusty circumstellar disk}.
\newblock {\em Nature}, 487:74, 2012.

\bibitem{Melosh1989}
H.~J. {Melosh} and A.~M. {Vickery}.
\newblock {Impact erosion of the primordial atmosphere of Mars}.
\newblock {\em Nature}, 338:487, 1989.

\bibitem{Meng2014}
Huan Y.~A. {Meng}, Kate Y.~L. {Su}, George~H. {Rieke}, David~J. {Stevenson},
  Peter {Plavchan}, Wiphu {Rujopakarn}, Carey~M. {Lisse}, Saran {Poshyachinda},
  and Daniel~E. {Reichart}.
\newblock {Large impacts around a solar-analog star in the era of terrestrial
  planet formation}.
\newblock {\em Science}, 345:1032, 2014.

\bibitem{Mennesson2014}
B.~{Mennesson}, R.~{Millan-Gabet}, E.~{Serabyn}, M.~M. {Colavita}, O.~{Absil},
  G.~{Bryden}, M.~{Wyatt}, W.~{Danchi}, D.~{Defr{\`e}re}, O.~{Dor{\'e}},
  P.~{Hinz}, M.~{Kuchner}, S.~{Ragland}, N.~{Scott}, K.~{Stapelfeldt},
  W.~{Traub}, and J.~{Woillez}.
\newblock {Constraining the Exozodiacal Luminosity Function of Main-sequence
  Stars: Complete Results from the Keck Nuller Mid-infrared Surveys}.
\newblock {\em ApJ}, 797:119, 2014.

\bibitem{Morbidelli2018}
A.~{Morbidelli}, D.~{Nesvorny}, V.~{Laurenz}, S.~{Marchi}, D.~C. {Rubie},
  L.~{Elkins-Tanton}, M.~{Wieczorek}, and S.~{Jacobson}.
\newblock {The timeline of the lunar bombardment: Revisited}.
\newblock {\em Icarus}, 305:262, 2018.

\bibitem{Nesvorny2010}
David {Nesvorn{\'y}}, Peter {Jenniskens}, Harold~F. {Levison}, William~F.
  {Bottke}, David {Vokrouhlick{\'y}}, and Matthieu {Gounelle}.
\newblock {Cometary Origin of the Zodiacal Cloud and Carbonaceous
  Micrometeorites. Implications for Hot Debris Disks}.
\newblock {\em ApJ}, 713:816, 2010.

\bibitem{Nguyen2010}
H.~T. {Nguyen}, B.~{Schulz}, L.~{Levenson}, A.~{Amblard}, V.~{Arumugam},
  H.~{Aussel}, T.~{Babbedge}, A.~{Blain}, J.~{Bock}, A.~{Boselli}, V.~{Buat},
  N.~{Castro-Rodriguez}, A.~{Cava}, P.~{Chanial}, E.~{Chapin}, D.~L.
  {Clements}, A.~{Conley}, L.~{Conversi}, A.~{Cooray}, C.~D. {Dowell},
  E.~{Dwek}, S.~{Eales}, D.~{Elbaz}, M.~{Fox}, A.~{Franceschini}, W.~{Gear},
  J.~{Glenn}, M.~{Griffin}, M.~{Halpern}, E.~{Hatziminaoglou}, E.~{Ibar},
  K.~{Isaak}, R.~J. {Ivison}, G.~{Lagache}, N.~{Lu}, S.~{Madden}, B.~{Maffei},
  G.~{Mainetti}, L.~{Marchetti}, G.~{Marsden}, J.~{Marshall}, B.~{O'Halloran},
  S.~J. {Oliver}, A.~{Omont}, M.~J. {Page}, P.~{Panuzzo}, A.~{Papageorgiou},
  C.~P. {Pearson}, I.~{Perez Fournon}, M.~{Pohlen}, N.~{Rangwala},
  D.~{Rigopoulou}, D.~{Rizzo}, I.~G. {Roseboom}, M.~{Rowan-Robinson},
  D.~{Scott}, N.~{Seymour}, D.~L. {Shupe}, A.~J. {Smith}, J.~A. {Stevens},
  M.~{Symeonidis}, M.~{Trichas}, K.~E. {Tugwell}, M.~{Vaccari},
  I.~{Valtchanov}, L.~{Vigroux}, L.~{Wang}, R.~{Ward}, D.~{Wiebe}, G.~{Wright},
  C.~K. {Xu}, and M.~{Zemcov}.
\newblock {HerMES: The SPIRE confusion limit}.
\newblock {\em A\&A}, 518:L5, 2010.

\bibitem{Olofsson2020}
J.~{Olofsson}, J.~{Milli}, A.~{Bayo}, Th. {Henning}, and N.~{Engler}.
\newblock {The challenge of measuring the phase function of debris discs.
  Application to HR 4796 A}.
\newblock {\em A\&A}, 640:A12, 2020.

\bibitem{Owen2019}
James~E. {Owen} and Juna~A. {Kollmeier}.
\newblock {Radiation pressure clear-out of dusty photoevaporating discs}.
\newblock {\em MNRAS}, 487:3702, 2019.

\bibitem{Patel2015}
B.~H.. {Patel}, C.~{Percivalle}, D.~J. {Ritson}, C.~D. {Duffy}, and J.~D.
  {Sutherland}.
\newblock {Common origins of RNA, protein and lipid precursors in a
  cyanosulfidic protometabolism}.
\newblock {\em Nature Chem}, 7:301, 2015.

\bibitem{Pearce2020}
Tim~D. {Pearce}, Alexander~V. {Krivov}, and Mark {Booth}.
\newblock {Gas trapping of hot dust around main-sequence stars}.
\newblock {\em MNRAS}, 498:2798, 2020.

\bibitem{Pearce2014}
Tim~D. {Pearce} and Mark~C. {Wyatt}.
\newblock {Dynamical evolution of an eccentric planet and a less massive debris
  disc}.
\newblock {\em MNRAS}, 443:2541, 2014.

\bibitem{Quanz2019}
Sascha~P. {Quanz}, Olivier {Absil}, Daniel {Angerhausen}, Willy {Benz}, Xavier
  {Bonfils}, Jean-Philippe {Berger}, Matteo {Brogi}, Juan {Cabrera}, William~C.
  {Danchi}, Denis {Defr{\`e}re}, Ewine {van Dishoeck}, David {Ehrenreich},
  Steve {Ertel}, Jonathan {Fortney}, Scott {Gaudi}, Julien {Girard}, Adrian
  {Glauser}, John~Lee {Grenfell}, Michael {Ireland}, Markus {Janson}, Jens
  {Kammerer}, Daniel {Kitzmann}, Stefan {Kraus}, Oliver {Krause}, Lucas
  {Labadie}, Sylvestre {Lacour}, Tim {Lichtenberg}, Michael {Line}, Hendrik
  {Linz}, J{\'e}r{\^o}me {Loicq}, Bertrand {Mennesson}, Michael~R. {Meyer},
  Yamila {Miguel}, John {Monnier}, Mamadou {N'Diaye}, Enric {Pall{\'e}}, Didier
  {Queloz}, Heike {Rauer}, Ignasi {Ribas}, Sarah {Rugheimer}, Franck {Selsis},
  Gene {Serabyn}, Ignas {Snellen}, Alessandro {Sozzetti}, Karl~R.
  {Stapelfeldt}, Amaury {Triaud}, St{\'e}phane {Udry}, and Mark {Wyatt}.
\newblock {Atmospheric characterization of terrestrial exoplanets in the
  mid-infrared: biosignatures, habitability \& diversity}.
\newblock {\em arXiv:1908.01316}, 2019.

\bibitem{Rieke2016}
G.~H. {Rieke}, Andr{\'a}s {G{\'a}sp{\'a}r}, and N.~P. {Ballering}.
\newblock {Magnetic Grain Trapping and the Hot Excesses around Early-type
  Stars}.
\newblock {\em ApJ}, 816:50, 2016.

\bibitem{Rieke2005}
G.~H. {Rieke}, K.~Y.~L. {Su}, J.~A. {Stansberry}, D.~{Trilling}, G.~{Bryden},
  J.~{Muzerolle}, B.~{White}, N.~{Gorlova}, E.~T. {Young}, C.~A. {Beichman},
  K.~R. {Stapelfeldt}, and D.~C. {Hines}.
\newblock {Decay of Planetary Debris Disks}.
\newblock {\em ApJ}, 620:1010, 2005.

\bibitem{Rigley2020}
Jessica~K. {Rigley} and Mark~C. {Wyatt}.
\newblock {Dust size and spatial distributions in debris discs: predictions for
  exozodiacal dust dragged in from an exo-Kuiper belt}.
\newblock {\em MNRAS}, 497:1143, 2020.

\bibitem{Roberge2012}
Aki {Roberge}, Christine~H. {Chen}, Rafael {Millan-Gabet}, Alycia~J.
  {Weinberger}, Philip~M. {Hinz}, Karl~R. {Stapelfeldt}, Olivier {Absil},
  Marc~J. {Kuchner}, and Geoffrey {Bryden}.
\newblock {The Exozodiacal Dust Problem for Direct Observations of Exo-Earths}.
\newblock {\em PASP}, 124:799, 2012.

\bibitem{Roelfsema2018}
P.~R. {Roelfsema}, H.~{Shibai}, L.~{Armus}, D.~{Arrazola}, M.~{Audard}, M.~D.
  {Audley}, C.~M. {Bradford}, I.~{Charles}, P.~{Dieleman}, Y.~{Doi},
  L.~{Duband}, M.~{Eggens}, J.~{Evers}, I.~{Funaki}, J.~R. {Gao}, M.~{Giard},
  A.~{di Giorgio}, L.~M. {Gonz{\'a}lez Fern{\'a}ndez}, M.~{Griffin}, F.~P.
  {Helmich}, R.~{Hijmering}, R.~{Huisman}, D.~{Ishihara}, N.~{Isobe},
  B.~{Jackson}, H.~{Jacobs}, W.~{Jellema}, I.~{Kamp}, H.~{Kaneda}, M.~{Kawada},
  F.~{Kemper}, F.~{Kerschbaum}, P.~{Khosropanah}, K.~{Kohno}, P.~P. {Kooijman},
  O.~{Krause}, J.~{van der Kuur}, J.~{Kwon}, W.~M. {Laauwen}, G.~{de Lange},
  B.~{Larsson}, D.~{van Loon}, S.~C. {Madden}, H.~{Matsuhara}, F.~{Najarro},
  T.~{Nakagawa}, D.~{Naylor}, H.~{Ogawa}, T.~{Onaka}, S.~{Oyabu},
  A.~{Poglitsch}, V.~{Reveret}, L.~{Rodriguez}, L.~{Spinoglio}, I.~{Sakon},
  Y.~{Sato}, K.~{Shinozaki}, R.~{Shipman}, H.~{Sugita}, T.~{Suzuki}, F.~F.~S.
  {van der Tak}, J.~{Torres Redondo}, T.~{Wada}, S.~Y. {Wang}, C.~K.
  {Wafelbakker}, H.~{van Weers}, S.~{Withington}, B.~{Vandenbussche},
  T.~{Yamada}, and I.~{Yamamura}.
\newblock {SPICA-A Large Cryogenic Infrared Space Telescope: Unveiling the
  Obscured Universe}.
\newblock {\em PASA}, 35:e030, 2018.

\bibitem{Schoonenberg2017}
Djoeke {Schoonenberg} and Chris~W. {Ormel}.
\newblock {Planetesimal formation near the snowline: in or out?}
\newblock {\em A\&A}, 602:A21, 2017.

\bibitem{Shannon2016}
Andrew {Shannon}, Amy {Bonsor}, Quentin {Kral}, and Elisabeth {Matthews}.
\newblock {The unseen planets of double belt debris disc systems}.
\newblock {\em MNRAS}, 462:L116, 2016.

\bibitem{Sibthorpe2018}
B.~{Sibthorpe}, G.~M. {Kennedy}, M.~C. {Wyatt}, J.~F. {Lestrade}, J.~S.
  {Greaves}, B.~C. {Matthews}, and G.~{Duch{\^e}ne}.
\newblock {Analysis of the Herschel DEBRIS Sun-like star sample}.
\newblock {\em MNRAS}, 475:3046, 2018.

\bibitem{Siegler2007}
Nick {Siegler}, James {Muzerolle}, Erick~T. {Young}, George~H. {Rieke}, Eric~E.
  {Mamajek}, David~E. {Trilling}, Nadya {Gorlova}, and Kate Y.~L. {Su}.
\newblock {Spitzer 24 {\ensuremath{\mu}}m Observations of Open Cluster IC 2391
  and Debris Disk Evolution of FGK Stars}.
\newblock {\em ApJ}, 654:580, 2007.

\bibitem{Smith1984}
B.~A. {Smith} and R.~J. {Terrile}.
\newblock {A Circumstellar Disk around {\ensuremath{\beta}} Pictoris}.
\newblock {\em Science}, 226:1421, 1984.

\bibitem{Su2019}
Kate Y.~L. {Su}, Alan~P. {Jackson}, Andr{\'a}s {G{\'a}sp{\'a}r}, George~H.
  {Rieke}, Ruobing {Dong}, Johan {Olofsson}, G.~M. {Kennedy}, Zo{\"e}~M.
  {Leinhardt}, Renu {Malhotra}, Michael {Hammer}, Huan Y.~A. {Meng},
  W.~{Rujopakarn}, Joseph~E. {Rodriguez}, Joshua {Pepper}, D.~E. {Reichart},
  David {James}, and Keivan~G. {Stassun}.
\newblock {Extreme Debris Disk Variability: Exploring the Diverse Outcomes of
  Large Asteroid Impacts During the Era of Terrestrial Planet Formation}.
\newblock {\em AJ}, 157:202, 2019.

\bibitem{Teague2018}
Richard {Teague}, Jaehan {Bae}, Edwin~A. {Bergin}, Tilman {Birnstiel}, and
  Daniel {Foreman-Mackey}.
\newblock {A Kinematical Detection of Two Embedded Jupiter-mass Planets in HD
  163296}.
\newblock {\em ApJL}, 860:L12, 2018.

\bibitem{Thureau2014}
N.~D. {Thureau}, J.~S. {Greaves}, B.~C. {Matthews}, G.~{Kennedy},
  N.~{Phillips}, M.~{Booth}, G.~{Duch{\^e}ne}, J.~{Horner}, D.~R. {Rodriguez},
  B.~{Sibthorpe}, and M.~C. {Wyatt}.
\newblock {An unbiased study of debris discs around A-type stars with
  Herschel}.
\newblock {\em MNRAS}, 445:2558, 2014.

\bibitem{Veras2016}
Dimitri {Veras}.
\newblock {Post-main-sequence planetary system evolution}.
\newblock {\em RSOS}, 3:150571, 2016.

\bibitem{Vitense2012}
Ch. {Vitense}, A.~V. {Krivov}, H.~{Kobayashi}, and T.~{L{\"o}hne}.
\newblock {An improved model of the Edgeworth-Kuiper debris disk}.
\newblock {\em A\&A}, 540:A30, 2012.

\bibitem{Williams2019}
Jonathan~P. {Williams}, Lucas {Cieza}, Antonio {Hales}, Megan {Ansdell}, Dary
  {Ruiz-Rodriguez}, Simon {Casassus}, Sebastian {Perez}, and Alice {Zurlo}.
\newblock {The Ophiuchus DIsk Survey Employing ALMA (ODISEA): Disk Dust Mass
  Distributions across Protostellar Evolutionary Classes}.
\newblock {\em ApJL}, 875:L9, 2019.

\bibitem{Wyatt2003}
M.~C. {Wyatt}.
\newblock {Resonant Trapping of Planetesimals by Planet Migration: Debris Disk
  Clumps and Vega's Similarity to the Solar System}.
\newblock {\em ApJ}, 598:1321, 2003.

\bibitem{Wyatt2005}
M.~C. {Wyatt}.
\newblock {Spiral structure when setting up pericentre glow: possible giant
  planets at hundreds of AU in the HD 141569 disk}.
\newblock {\em A\&A}, 440:937, 2005.

\bibitem{Wyatt2008}
M.~C. {Wyatt}.
\newblock {Evolution of debris disks.}
\newblock {\em ARA\&A}, 46:339, 2008.

\bibitem{Wyatt2017}
M.~C. {Wyatt}, A.~{Bonsor}, A.~P. {Jackson}, S.~{Marino}, and A.~{Shannon}.
\newblock {How to design a planetary system for different scattering outcomes:
  giant impact sweet spot, maximizing exocomets, scattered discs}.
\newblock {\em MNRAS}, 464:3385, 2017.

\bibitem{Wyatt1999}
M.~C. {Wyatt}, S.~F. {Dermott}, C.~M. {Telesco}, R.~S. {Fisher}, K.~{Grogan},
  E.~K. {Holmes}, and R.~K. {Pi{\~n}a}.
\newblock {How Observations of Circumstellar Disk Asymmetries Can Reveal Hidden
  Planets: Pericenter Glow and Its Application to the HR 4796 Disk}.
\newblock {\em ApJ}, 527:918, 1999.

\bibitem{Wyatt2020}
M.~C. {Wyatt}, Q.~{Kral}, and C.~A. {Sinclair}.
\newblock {Susceptibility of planetary atmospheres to mass-loss and growth by
  planetesimal impacts: the impact shoreline}.
\newblock {\em MNRAS}, 491:782, 2020.

\bibitem{Wyatt2015}
M.~C. {Wyatt}, O.~{Pani{\'c}}, G.~M. {Kennedy}, and L.~{Matr{\`a}}.
\newblock {Five steps in the evolution from protoplanetary to debris disk}.
\newblock {\em Ap\&SS}, 357:103, 2015.

\bibitem{Wyatt2007}
M.~C. {Wyatt}, R.~{Smith}, K.~Y.~L. {Su}, G.~H. {Rieke}, J.~S. {Greaves}, C.~A.
  {Beichman}, and G.~{Bryden}.
\newblock {Steady State Evolution of Debris Disks around A Stars}.
\newblock {\em ApJ}, 663:365, 2007.

\bibitem{Wyatt2016}
Mark~C. {Wyatt} and Alan~P. {Jackson}.
\newblock {Insights into Planet Formation from Debris Disks. II. Giant Impacts
  in Extrasolar Planetary Systems}.
\newblock {\em SSRv}, 205:231, 2016.

\end{thebibliography}
\end{document}